\documentclass[useAMS,usenatbib]{mn2e}
\usepackage{subfigure}
\usepackage{graphicx}

\title[Dusty torus signatures in the WLRG PKS 0043-42]{Clear detection of dusty torus signatures in a Weak-Line Radio Galaxy: the case of PKS 0043-42}
\author[C. Ramos Almeida et al.]
{\parbox{\textwidth}{C. Ramos Almeida$^{1}$\thanks{E-mail:C.Ramos@sheffield.ac.uk},
D. Dicken$^{2}$,
C. Tadhunter$^{1}$,
A. Asensio Ramos$^{3,4}$,
K. J. Inskip$^{5}$,
M. J. Hardcastle$^{6}$,
B. Mingo$^{6}$
}\vspace{0.4cm}\\
\parbox{\textwidth}{$^{1}$Department of Physics and Astronomy, University of Sheffield, Sheffield, S3 7RH, UK\\
$^{2}$Department of Physics and Astronomy, Rochester Institute of Technology, 84 Lomb Memorial Drive, Rochester NY 14623, USA\\
$^{3}$Instituto de Astrof\'\i sica de Canarias (IAC), C/V\'\i a L\'{a}ctea, s/n, E-38205, La Laguna, Tenerife, Spain\\
$^{4}$Departamento de Astrof\' isica, Universidad de La Laguna, E-38205, La Laguna, Tenerife, Spain\\
$^{5}$Max-Planck-Institut f\"ur Astronomie, K\"oningstuhl 17, D-69117 Heidelberg, Germany\\
$^{6}$School of Physics, Astronomy and Mathematics, University of Hertfordshire, College Lane, Hatfield, Hertfordshire AL10 9AB, UK}
}
\begin{document}

\date{}

\pagerange{\pageref{firstpage}--\pageref{lastpage}} \pubyear{2010}

\maketitle

\label{firstpage}

\begin{abstract}
We report the clearest detection to date of dusty torus signatures in a Weak-Line Radio Galaxy (WLRG). 
The deep Spitzer InfraRed Spectrograph (IRS) rest-frame mid-infrared (MIR) spectrum of the WLRG PKS 0043-42 (z=0.116) shows 
a clear spectral turnover at $\lambda\ga$ 20 \micron~suggestive of warm dust, as well as a 9.7 \micron~silicate absorption feature.
In addition, the hard X-ray results, based on Chandra data, strongly support a
picture in which PKS 0043$-$42 has a torus and accretion disc more typical of Strong-Line Radio Galaxies (SLRGs).
The MIR and X-ray spectra are markedly different from those of other WLRGs at similar redshifts, and here we show 
that the former can be successfully fitted with clumpy torus models with parameters characteristic of 
Type-2 AGN tori: close to edge-on ($i$=74\degr) and relatively broad ($\sigma$=60\degr), with an outer radius of 2 pc, 
N$_H$=1.6$\pm^{0.2}_{0.1}\times$10$^{23}~cm^{-2}$, and AGN bolometric luminosity L$_{bol}^{AGN}$ = 1.6$\pm^{0.2}_{0.1}\times$10$^{44}~erg~s^{-1}$.  
The presence of a compact torus in PKS 0043-42 provides evidence that this WLRG is fuelled by cold, rather than hot, 
gas accretion. 
We suggest that WLRGs are a diverse population, and PKS 0043-42 may represent a type of radio galaxy in which 
the AGN activity has been recently re-triggered as a consequence of intermittent gas supply, or in which the
covering factor of the Narrow-Line Region (NLR) clouds is relatively low.
\end{abstract}


\section{Introduction}
\label{intro}

Most powerful radio galaxies (PRGs) can be classified as Narrow- and Broad-Line 
Radio Galaxies/quasars (NLRGs, BLRGs/QSOs) on the basis of their spectral features. The
unified model for active galactic nuclei (AGN; \citealt{Antonucci93,Urry95}), 
proposes the existence of a pc-scale obscuring toroidal structure to account for the observed 
differences between their spectra. 
However, there is also a third class, the WLRGs (also known as Low-Excitation Galaxies; LEGs), whose optical spectra are dominated 
by the stellar continua of the host galaxies \citep{Laing94,Tadhunter98} without the prominent emission 
lines characteristic of NLRGs and BLRGs (which can be grouped as SLRGs, or alternatively, High-Excitation Galaxies; HEGs).

The nature of the physical parameters which lead to the division between SLRGs and WLRGs
still remains unclear. 
Previous studies failed to explain the differences between the two types as due to different 
or time-varying accretion rates \citep{Ghisellini01}. 
An alternative interpretation is that SLRGs are powered by cold gas accretion, while WLRGs are fuelled by 
accretion of hot gas provided by the reservoir 
of their X-ray gaseous coronae \citep{Allen06,Best06,Hardcastle07,Balmaverde08,Buttiglione10}.
The high temperatures of the hot gas would prevent the formation of the ``cold'' structures 
(e.g. the Broad-Line Region and the torus). 
This hypothesis explains the fact that WLRGs do not normally show the high level of X-ray absorption expected 
from a molecular torus \citep{Hardcastle09}.

SLRGs are almost invariably associated with Fanaroff-Riley II sources (FRII). On the other 
hand, while most WLRGs show Fanaroff-Riley I (FRI) radio morphologies, some are classified 
as FRIIs \citep{Laing94}. 
Evidence for nuclear warm dust emission has been found in the MIR
for FRII/SLRGs \citep{Ogle06,Wolk10} and, to a lesser extent, 
for a few FRI and FRII WLRGs \citep{Leipski09,Wolk10}.
However, none of the WLRGs so far studied have provided definitive evidence for a warm compact 
obscuring region e.g. in the form of a spectral turnover at $\sim$20 \micron~and a clear silicate absorption feature. 
This lack of conclusive detection of torus signatures in previous studies may be due to  
contamination of the MIR data by stellar photospheric and starburst-heated dust components, 
which generally produces a broad minimum in the 8-10 \micron~range and can be confused with a silicate absorption feature.
Strong PAH (in particular the 8.2 and 11.3 \micron~features) and 
non-thermal emission from the synchrotron-emitting core sources can also mask the torus signatures. 
Based on extrapolation of the radio core data, the latter emission appears 
particularly important in WLRGs at MIR wavelengths \citep{Dicken08,Wolk10}.
All of these diluting components need to be accurately subtracted before the presence of warm dust can be deduced.

At optical wavelengths the galaxy PKS 0043-42 is classified as a WLRG at redshift z=0.116 \citep{Tadhunter93}, 
whereas at radio wavelengths it shows a clear FRII morphology \citep{Morganti93,Morganti99}. 
According to \citet{Tadhunter98}, the WLRGs are defined as a group of galaxies with [O III]$\lambda$5007 
emission line equivalent widths below 10\AA. PKS 0043-42 not only fullfills this criterion, but 
also has an [O III]$\lambda$5007 luminosity a factor of 10 lower than those of the general population of SLRGs at similar redshifts
and radio powers (L$_{[O III]}$=5$\times$10$^{40}$ erg~s$^{-1}$; \citealt{Tadhunter98,Dicken09}), as well as
line ratios indicating a low ionization state (e.g., [O III]$\lambda$5007/[O II]$\lambda$3727 $<$ 1; \citealt{Tadhunter98,Lewis03}).
The galaxy is at the center of a group or cluster, and from NIR imaging, \citet{Inskip10} reported a central isophotal twist 
and an excess of emission along the NS direction. They claimed that both are likely to be associated either with the presence of a 
dust lane, or with the apparent interaction with the companion object $\sim$15~kpc to the North.
However, the deep optical image presented in \citet{Ramos10} does not reveal any dust features in PKS 0043-42, but appears instead to confirm
the presence of the bridge with the companion galaxy \citep{Ramos10}.

In this study we report the detection of torus signatures in the MIR and X-ray spectra of the WLRG PKS 0043-42.
In the following sections we describe the observations, the SED modelling, the main results 
and their implications.
Throughhout this paper we assume a cosmology with H$_0$=73 km~s$^{-1}$~Mpc$^{-1}$; $\Omega_m$=0.27, and $\Omega_{\Lambda}$=0.73, 
which provides a luminosity distance of 516 Mpc for PKS 0043-42 (from the NASA/IPAC Extragalactic Database; NED).

\section{Observations}

\subsection{Spitzer observations}
\label{observations}

\begin{figure}
\centering
\includegraphics[width=10cm]{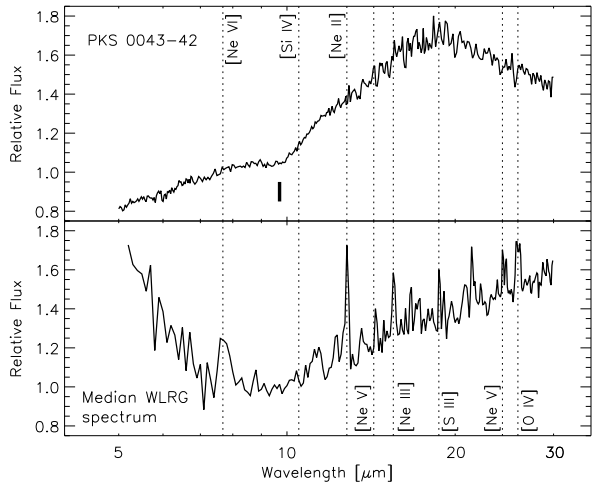}
\caption{Low-resolution IRS rest-frame spectrum of PKS 0043-42 in relative F$_{\nu}$ units (top). Note the emission bump peaking at 
$\sim$20 \micron~and the 9.7 \micron~silicate feature in absorption (short vertical line), both characteristic of a dusty torus.
The spectral features and shape are different to those of the 2Jy median WLRG spectrum (bottom). 
Typical AGN emission lines are labelled.}
\label{pks0043_spitzer}
\end{figure}

MIR spectroscopic observations of PKS 0043-42 were obtained using the IRS instrument \citep{Houck04} on  
the Spitzer satellite on 2008 Dec 12 under the program 2JYSPEC/50558 (PI: C. Tadhunter).
The observations were made in staring mode using the two low-resolution (R$\sim$60-120) IRS modules: 
the short-low (SL; 5.2--14.5 $\micron$) and the long-low (LL; 14-38 $\micron$). The slit widths are 
3.7\arcsec~and 10.7\arcsec~for the SL and LL modules respectively.
Four cycles of 60 s exposures were taken in each of the two SL modules, and two cycles of 120 s for the LL modules.  
The data were downloaded from the Spitzer archive in basic calibrated data (BCD) product form, 
processed with the S18.7.0 pipeline, and 
reduced and extracted using 
the {\sc smart} v.8.1.2 program developed by the IRS Team at Cornell University \citep{Higdon04,Lebouteiller10}. 
In extracting the fluxes we used the optimal extraction function which uses a super sampled PSF 
and weights the extracted spectra using the signal-to-noise of each pixel \citep{Lebouteiller10}. 
Note that the IRS pipeline automatically accounts for variable slit losses with wavelength. 
For each cycle the nod positions were subtracted in order to perform 
background subtraction on the image, as well as to counteract intrinsic sensitivity variations 
across the detector. For a more detailed description of the observations and
data reduction see Dicken et al., in prep. The two resulting spectra were then combined to 
produce the final spectrum shown in Figure \ref{pks0043_spitzer}. 
The absolute uncertainty in the flux calibration at the 1$\sigma$ level is $\sim$15\% for the SL and LL data, 
whereas for the spectral shape it is much less than that \citep{Decin04,Roellig04}. We have scaled the observed IRS 
spectrum to the Multiband Imaging Photometer for Spitzer (MIPS) 24 \micron~flux (11.1$\pm$0.2 mJy; \citealt{Dicken08}),
obtained using a 15\arcsec~aperture.
The discrepancy between the IRS and MIPS fluxes at 24 \micron~is only $\sim$10\%.


\begin{figure}
\centering
\subfigure[]{\includegraphics[width=4cm]{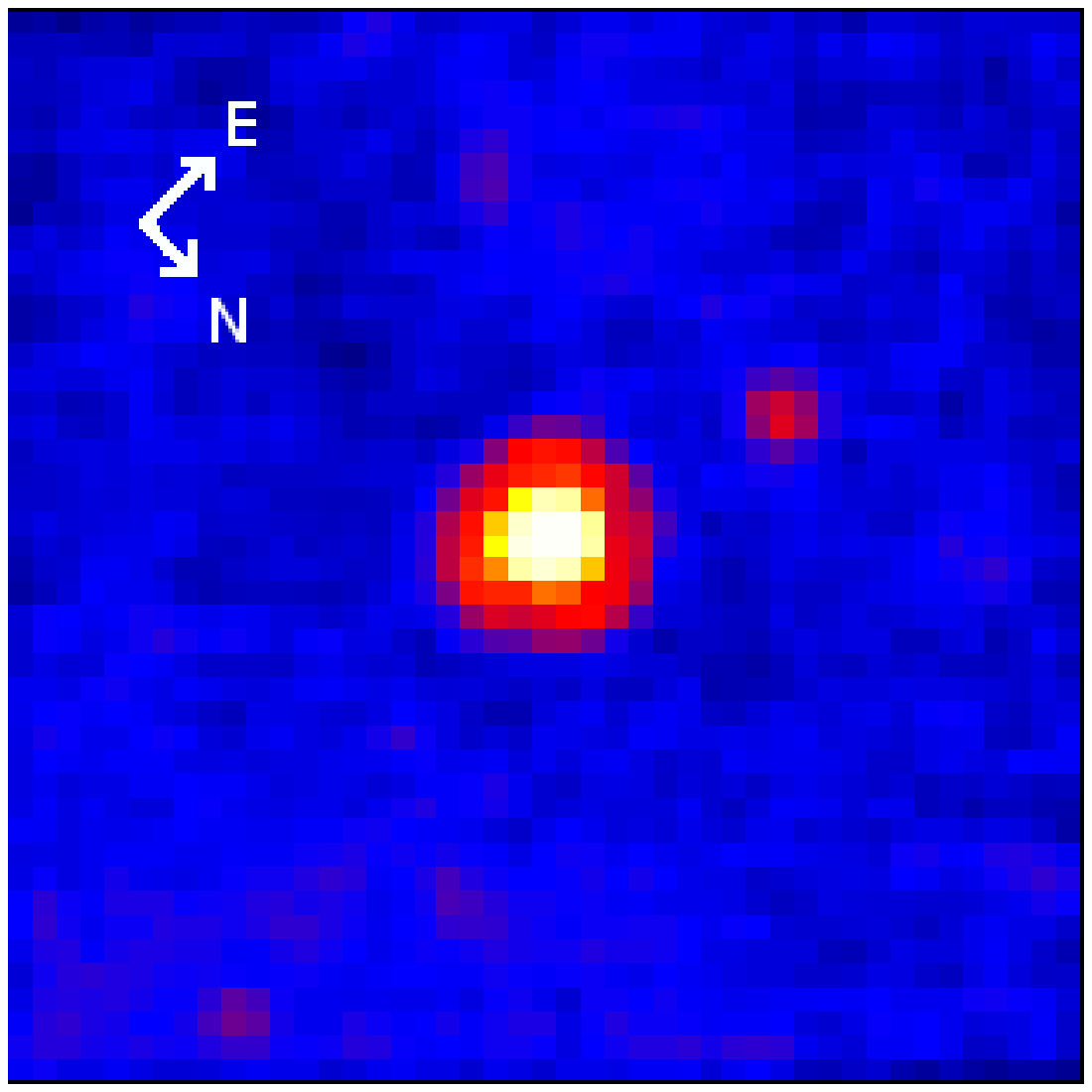}
\label{mips24}}
\subfigure[]{\includegraphics[width=4cm]{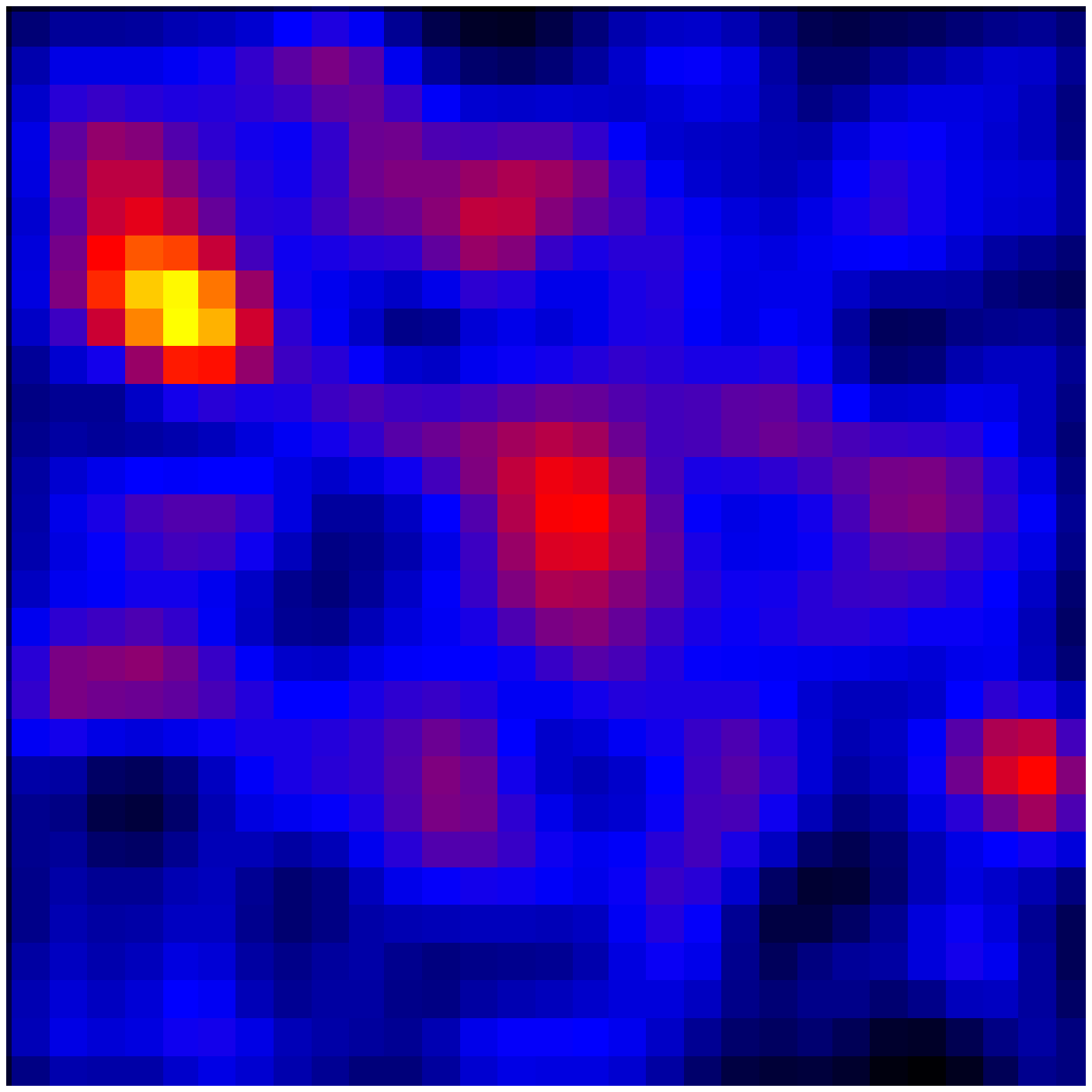}
\label{mips70}}
\caption{MIPS 24 \micron~(a) and 70 \micron~(b) images (the latter smoothed using a box size of 2 pixels) of PKS 0043-42.
Panels are 110\arcsec~side.}
\label{mips}
\end{figure}

In addition to the IRS spectrum, we compiled: 

\begin{enumerate}
\renewcommand{\theenumi}{(\arabic{enumi})}

\item A NIR upper limit flux of 0.45$\pm$0.02 mJy
derived from the GALFIT analysis of Ks-band SofI images presented in \citet{Inskip10}.
This value corresponds to an unresolved component of 0.93\arcsec~of FWHM.

\item A MIPS 70 \micron~flux of 9.9$\pm$3.0 mJy from \citet{Dicken08}. For the analysis performed here, 
we consider this flux an upper limit, because at this wavelength and with the aperture employed for 
calculating it (24\arcsec; \citealt{Dicken08}) the emission is likely to be dominated by dust heated by 
stars and/or AGN heated dust in the NLR. In Figure \ref{mips} we present the MIPS images of PKS 0043-42 
at 24 and 70 \micron~(from the 2Jy webpage: {\it http://2jy.extragalactic.info} and described in \citealt{Dicken08}),
to show the diference in the signal-to-noise level of the two detections.

\item High angular resolution radio core fluxes of 4.5$\pm$0.8 mJy at 2.3 GHz (13 cm) from \citet{Morganti97}, 
and 1.4$\pm$0.3 mJy at $\sim$15 GHz (2 cm) from \citet{Dicken08}.

\end{enumerate}

\begin{figure}
\centering
\includegraphics[width=9cm]{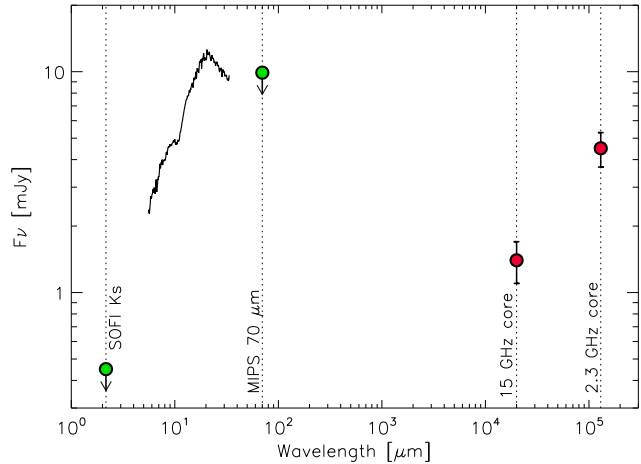}
\caption{SED of PKS 0043-42 including the NIR Ks-band upper limit, the IRS observed 
spectrum scaled to the MIPS 24 \micron~point, the MIPS 70 \micron~upper limit, and the 15 and 2.3 GHz radio core fluxes. The latter
indicate that non-thermal emission from the radio core does not contribute significantly to the MIR emission.}
\label{pks0043_radio}
\end{figure}

The IR-to-radio SED of PKS 0043-42 is shown in Figure \ref{pks0043_radio}. 
From the radio core data points 
it becomes clear that, for any reasonable extrapolation of the radio core spectrum, non-thermal emission 
does not contribute significantly to the MIR emission of 
this galaxy. By looking at the high resolution radio maps shown in \citet{Morganti97} we can also rule out
significant contamination (by extended emission from the radio lobes) of the IRS data  
(maximum slit width of 4.46\arcsec).

The MIR spectrum of PKS 0043-42 (Figure \ref{pks0043_spitzer}) shows a clear 9.7 \micron~silicate absorption feature 
and a spectral turnover at $\lambda\ga20$ \micron. These are the most convincing detections of torus signatures 
to date in a WLRG. We can quantify the 9.7 \micron~silicate feature in terms of its
strength, defined as $\tau_{9.7}^{app}$ = ln(F$_{cont}$)-ln(F$_{core}$). We calculate F$_{cont}$
using a cubic spline interpolation in logarithmic space over two intervals at shorter and longer
wavelength than the region covered by the silicate feature to fit the continuum and 
F$_{core}$ allowing the peak wavelength to vary, following \citet{Sirocky08}.
By doing this, we find $\tau_{9.7}^{app}$=0.2. The lack of strong emission lines in the IRS spectrum is also remarkable.

All of these features, as well as the spectral shape, are markedly
different from those of the median spectrum shown in the bottom panel of Figure \ref{pks0043_spitzer},
which was constructed using IRS data for all the WLRGs in the 2Jy sample 
with redshift 0.05$<$z$<$0.15 (excluding PKS 0034-01, which is affected by saturation problems), and
is representative of WLRGs at similar redshifts to PKS 0043-42 (see also \citealt{Leipski09}).
At short wavelengths, this average WLRG spectrum shows a blue spectral slope, which corresponds
to the tail of the photospheric emission from the stars in the host galaxy. At longer wavelengths, the continuum appears
relatively flat, although rising slowly with wavelength. This shape is likely produced by a combination 
of non-thermal emission and extended dust emission heated by stars \citep{Leipski09}. As mentioned in Section \ref{intro}, 
there is evidence for strong synchrotron contamination at MIR and far-infrared wavelengths in many WLRGs, based
on their IR-to-radio SEDs \citep{Dicken08,Wolk10}. 
Finally, emission lines such as [Ne II]$\lambda$12.8, [Ne III]$\lambda$15.5, and 
[O IV]$\lambda$25.9 \micron~are also detected in the median WLRG spectrum. 

\subsection{Chandra observations}
\label{chandra}

\begin{figure}
\centering
\includegraphics[width=6cm,angle=-90]{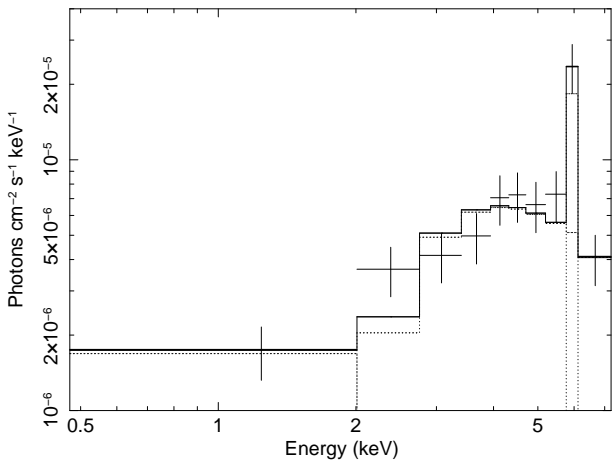}
\caption{{\it Chandra} ACIS-S spectrum of PKS 0043-42. The model consists of local absorption, 
intrinsic absorption, two power-laws and a Gaussian to account for the Fe emission line. 
The presence of this emission line is associated with accreting structures, 
which are typically seen in NLRGs, but have not been unequivocally observed in WLRGs before.}
\label{chandra_spec}
\end{figure}

PKS 0043$-$42 was observed with Chandra on 2009 Feb 05 (obsid
10319) as part of a systematic programme of Chandra observations
of the complete steep-spectrum 2Jy sample in the redshift range $0.05
< z < 0.2$ (details of all the observations, including further results
from the observation of PKS 0043$-$42, will be presented by Mingo et
al.\ in prep). The time on source was 18.4 ks. X-ray emission from the
nucleus was clearly detected, at the level of $\sim 150$ counts in the
0.5-5.0 keV energy band, and we extracted an X-ray spectrum of the
nucleus in the manner described by \citet{Hardcastle06}. This spectrum
is shown in Figure \ref{chandra_spec}.

The spectrum is clearly different from what is usually seen in WLRGs \citep{Hardcastle06},
namely a single power law with little or no intrinsic absorption: a
peak at high energies indicates the need for an absorbed component in
the model. Therefore, we fitted the standard model for narrow-line
radio galaxies, consisting of two power laws, one with absorption
fixed at the Galactic value ($N_{\rm H} = 2.7 \times 10^{20}$
cm$^{-2}$), and one with an additional free variable absorption.
Because of the small number of counts in the spectrum, we fixed the
photon indices of the unabsorbed and absorbed power laws to the values
used by \citet{Hardcastle06}, i.e. 2.0 and 1.7 respectively. A fit
of this model showed strong residuals at 6-7 keV, which could not be
removed by allowing the photon index of the absorbed power law to
vary; a good fit could only be obtained with the addition of a narrow
Gaussian feature with $E = 6.47^{+0.39}_{-0.04}$ keV, consistent with
being an Fe K$\alpha$ line (indicative of the presence of an accretion disk). 
The best-fitting absorption column density in this
model was $1.20 ^{+0.55}_{-0.36} \times 10^{23}$ cm$^{-2}$ and the
unabsorbed 2--10 keV luminosity of the heavily-absorbed power-law component was
$1.83 ^{+0.10}_{-0.12} \times 10^{43}$ erg s$^{-1}$. The high
obscuring column density and the clearly detected iron feature are both very
commonly seen in NLRGs, but have not so far been seen in bona fide
WLRGs (e.g. \citealt{Hardcastle09}). 

In addition, the X-ray luminosity
lies in the region populated by NLRGs on a plot of absorbed X-ray
luminosity against low-frequency radio luminosity (Fig.\ 3 of
\citealt{Hardcastle09}), as well as on the correlation between the 15 \micron~luminosity 
(5.8$\times$10$^{43}$ erg~s$^{-1}$, as measured from the IRS spectrum) and the unabsorbed 
X-ray luminosity (Fig.\ 10 of \citealt{Hardcastle09}). 
Thus, the X-ray results strongly support the
picture in which PKS 0043$-$42, while spectroscopically classified as a
WLRG, has a torus and accretion disc more typical of a NLRG.

\section{SED modelling}
\label{modelling}

Our aim is to investigate whether or not the IR SED of PKS 0043-42 can be reproduced
by torus models. 
The IR range (and particularly the MIR; 5-30 \micron) is key for setting constraints on the torus models, 
since the reprocessed radiation from the dust in the torus is re-emitted in this range.
Recent ground-based MIR imaging (see e.g. \citealt{Packham05,Radomski08}) and interferometric 
observations of nearby AGN (e.g. \citealt{Jaffe04,Tristram07}) reveal that the torus size is likely 
restricted to a few parsecs -- much smaller than the resolution of the IRS spectra. 
Thus, large aperture data such as those employed here may potentially be contaminated by
i) photospheric emission from stars, ii) extended dust emission heated by stars and/or the AGN, and iii) non-thermal emission. 
First, the optical spectrum and the lack of polycyclic aromatic hydrocarbons (PAHs) in the MIR 
suggest that star formation is not important in PKS 0043-42. Second, extended dust emission from the NLR
cannot be a substantial source of contamination either if we consider the low emission line luminosity 
measured for this galaxy \citep{Tadhunter98,Dicken09}.
Finally, in Section \ref{observations} we have shown that 
the non-thermal emission from the radio core and the lobes does not contribute significantly to the MIR emission of this object. 
Thus, the dominant contribution to our IR data must be nuclear dust heated by the AGN.


The clumpy dusty torus models of \citet{Nenkova02} hold that the dust
surrounding the central engine of an AGN is distributed in clumps, instead of homogeneously filling the torus volume.  
These clumps are distributed with a radial extent $Y = R_{o}/R_{d}$, where 
$R_{o}$ and $R_{d}$ are the outer and inner radius of the toroidal distribution, respectively. 
The clumpy database now contains 5$\times$10$^6$ models, calculated for a fine grid of model
parameters. 
To take into account the inherent degeneracy between these parameters when fitting observables,
here we use an updated version of the Bayesian inference tool {\it BayesClumpy}, developed 
by \citet{Asensio09}. {\it BayesClumpy} applies interpolation methods that are able to derive models for different 
combinations of the clumpy model parameters even if they are not present in the original database. 
For an example of the use of clumpy model fitting to IR SEDs using BayesClumpy, see \citet{Ramos09}.
The new version of {\it BayesClumpy} allows to fit photometric and spectroscopic data simultaneously.
Thus, here we use an interpolated version of the clumpy models of \citet{Nenkova08a,Nenkova08b} 
including the corrections for the previously erroneous AGN scaling factor (see erratum by 
\citealt{Nenkova10}).

\begin{figure*}
\centering
\par{
\includegraphics[width=5.6cm]{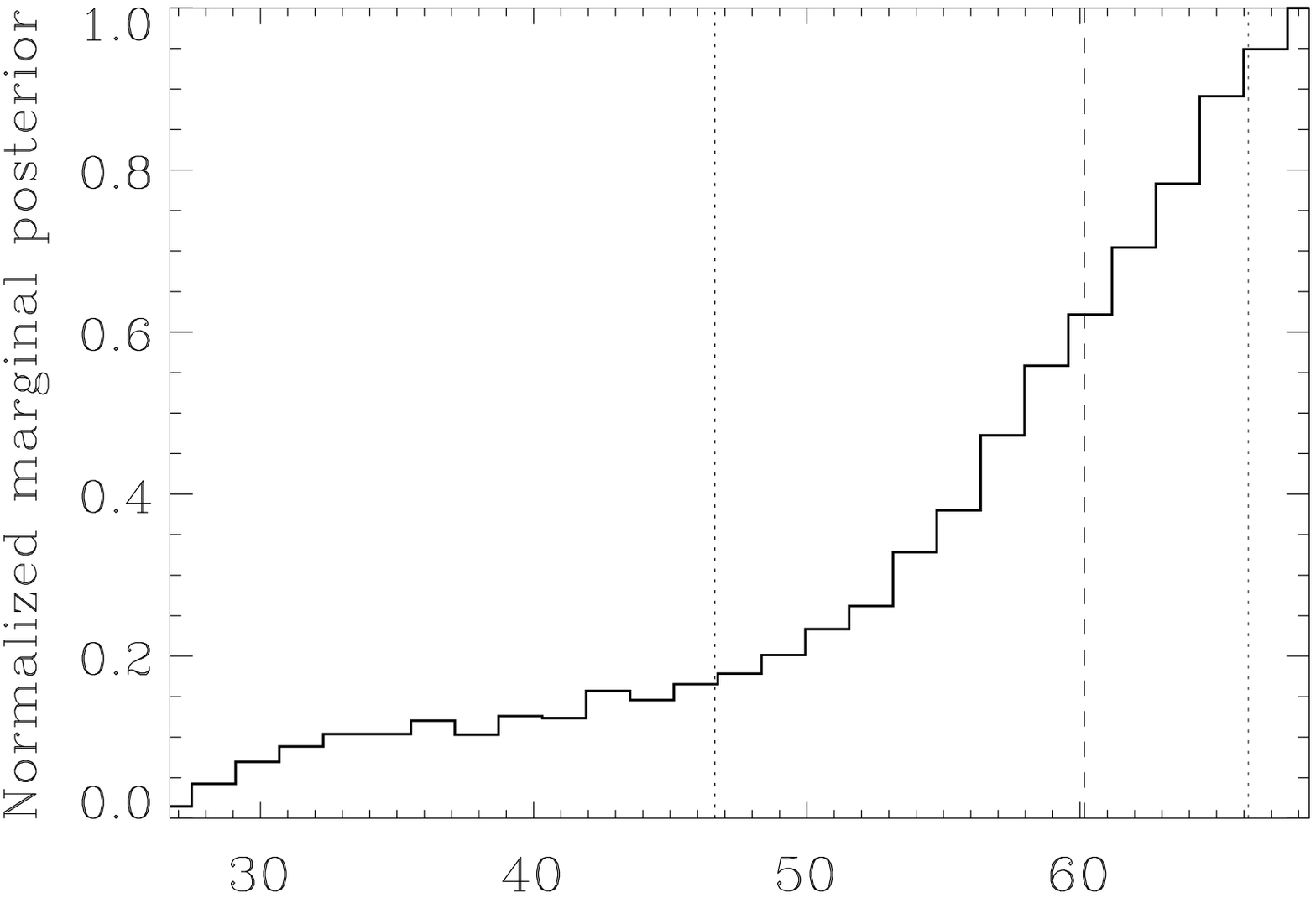}
\includegraphics[width=5.6cm]{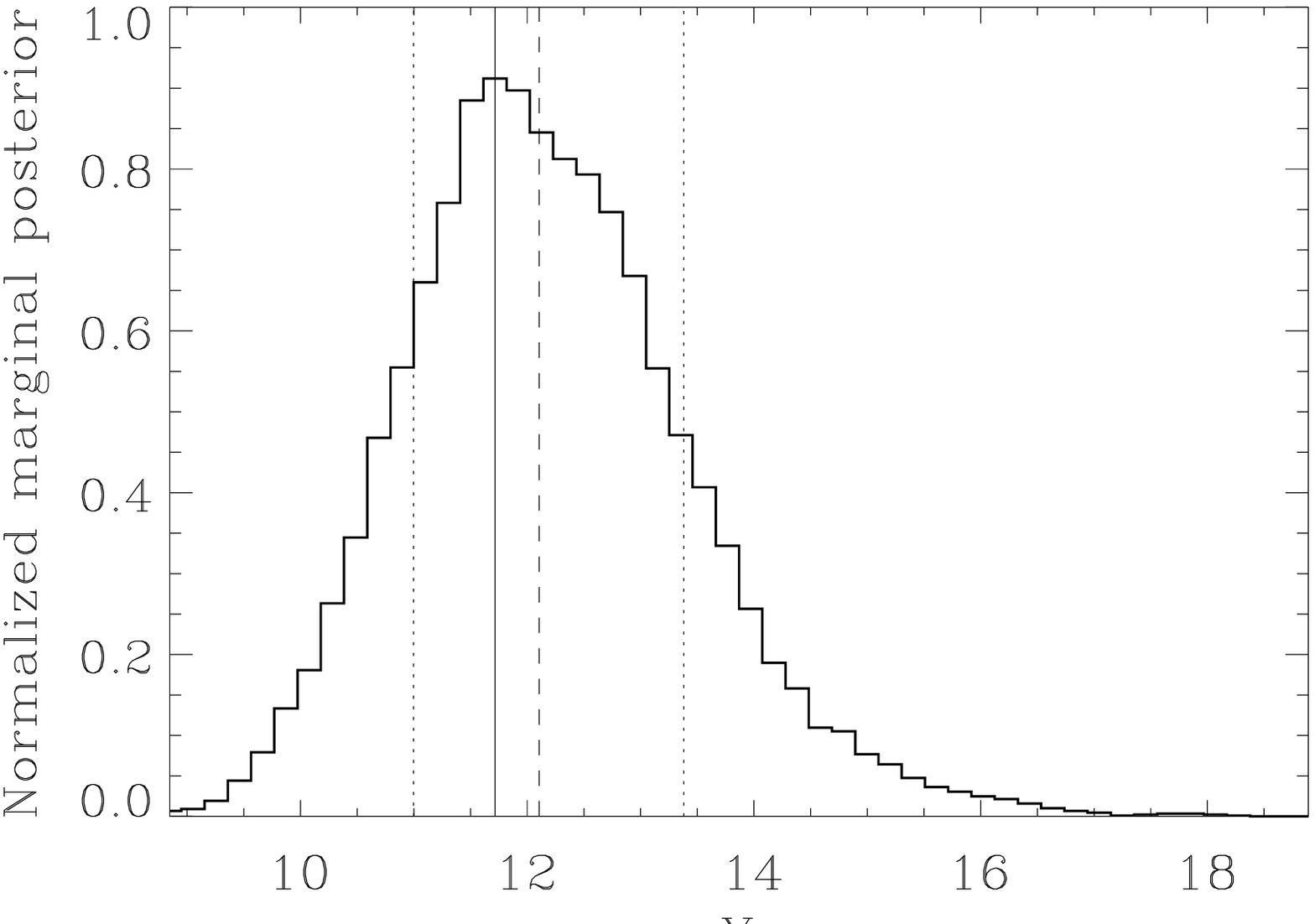}
\includegraphics[width=5.6cm]{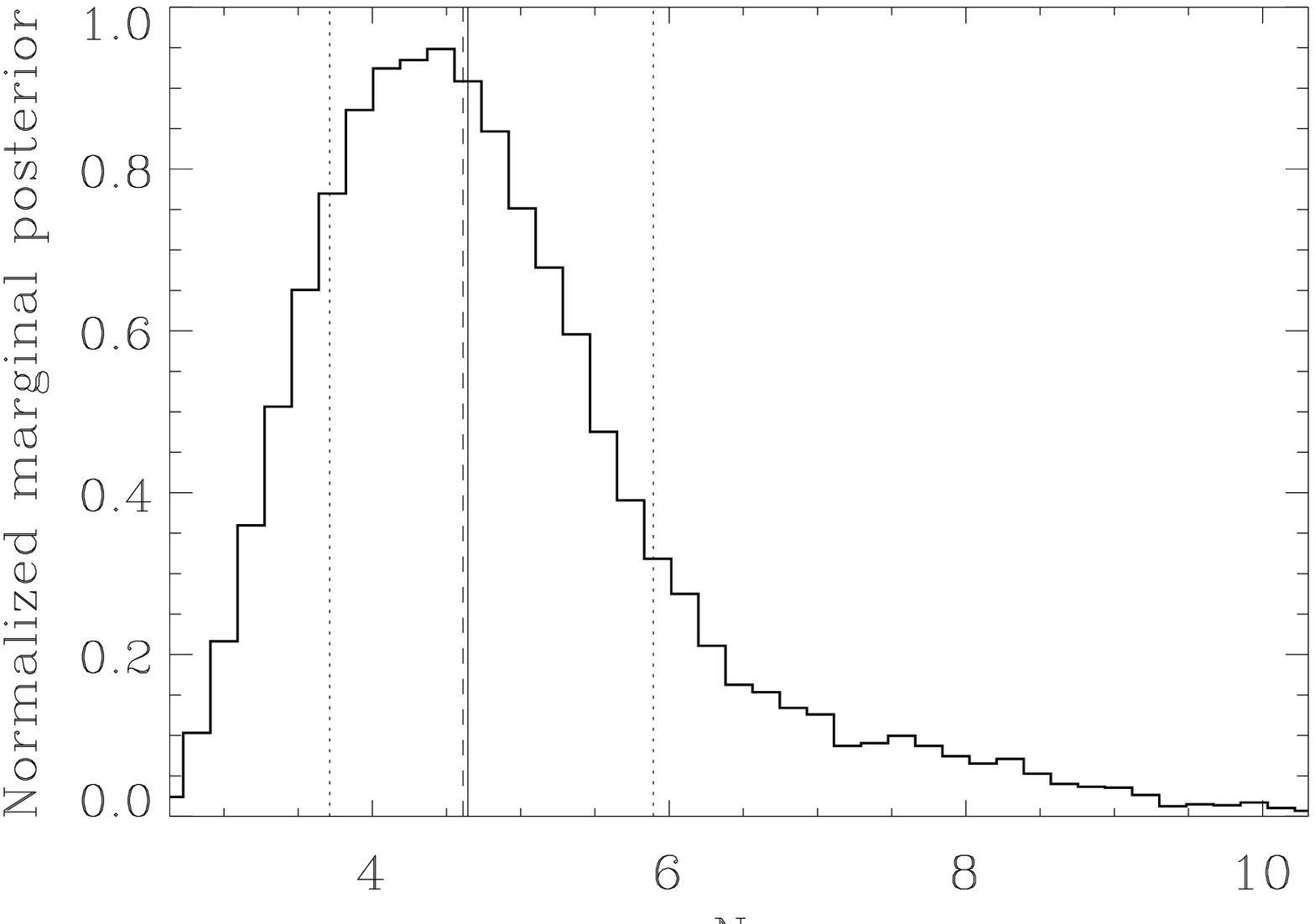}
\includegraphics[width=5.6cm]{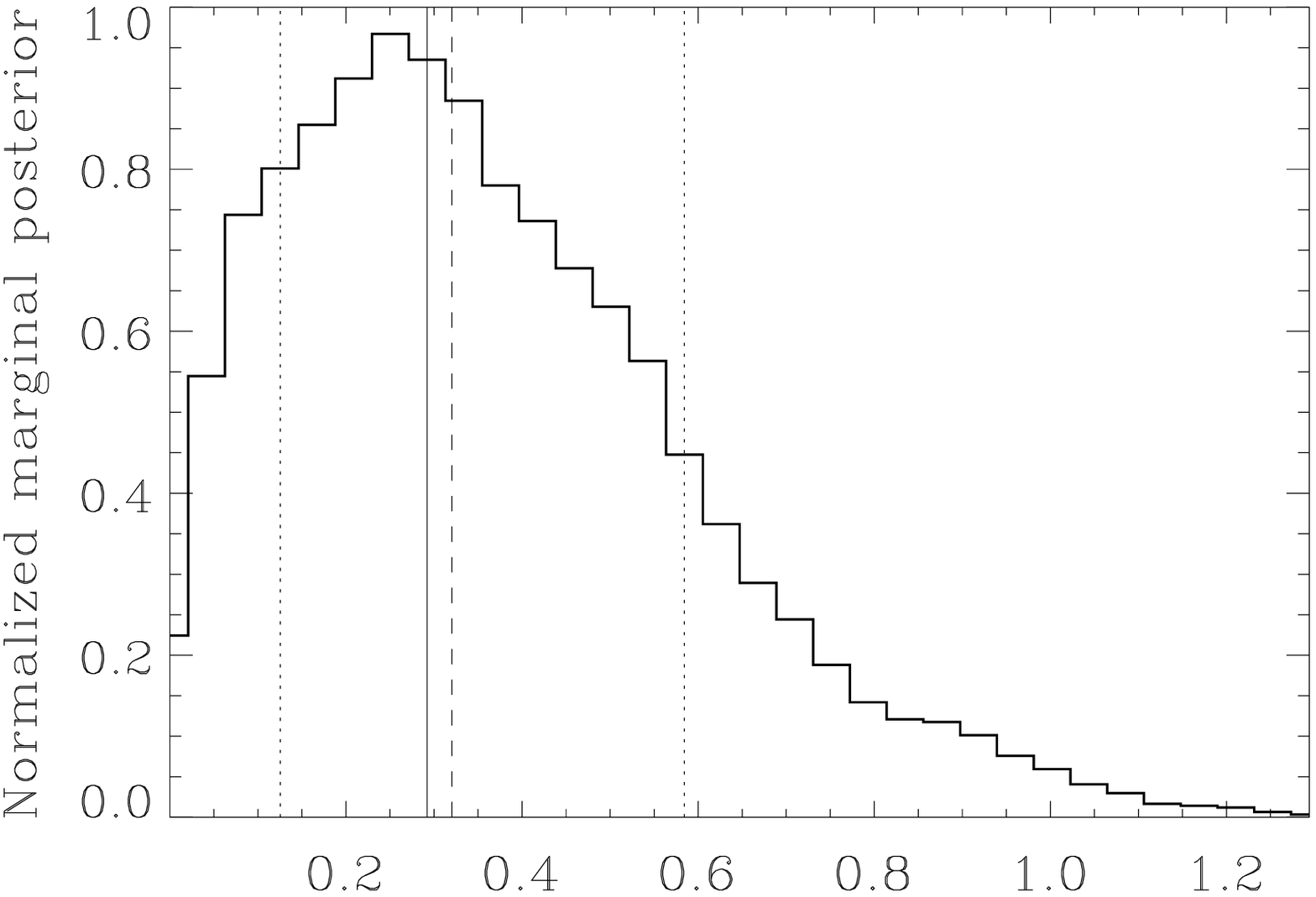}
\includegraphics[width=5.6cm]{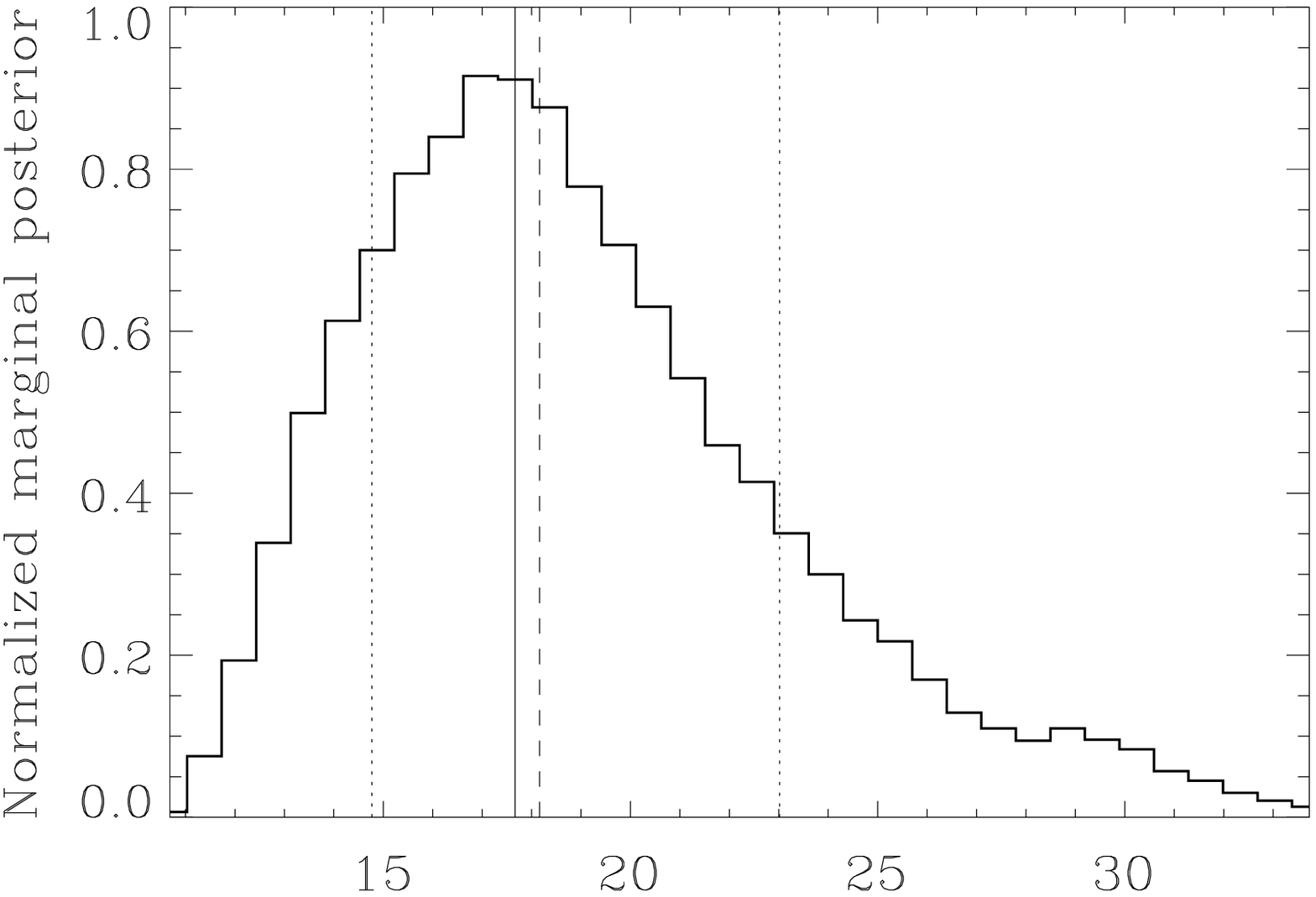}
\includegraphics[width=5.6cm]{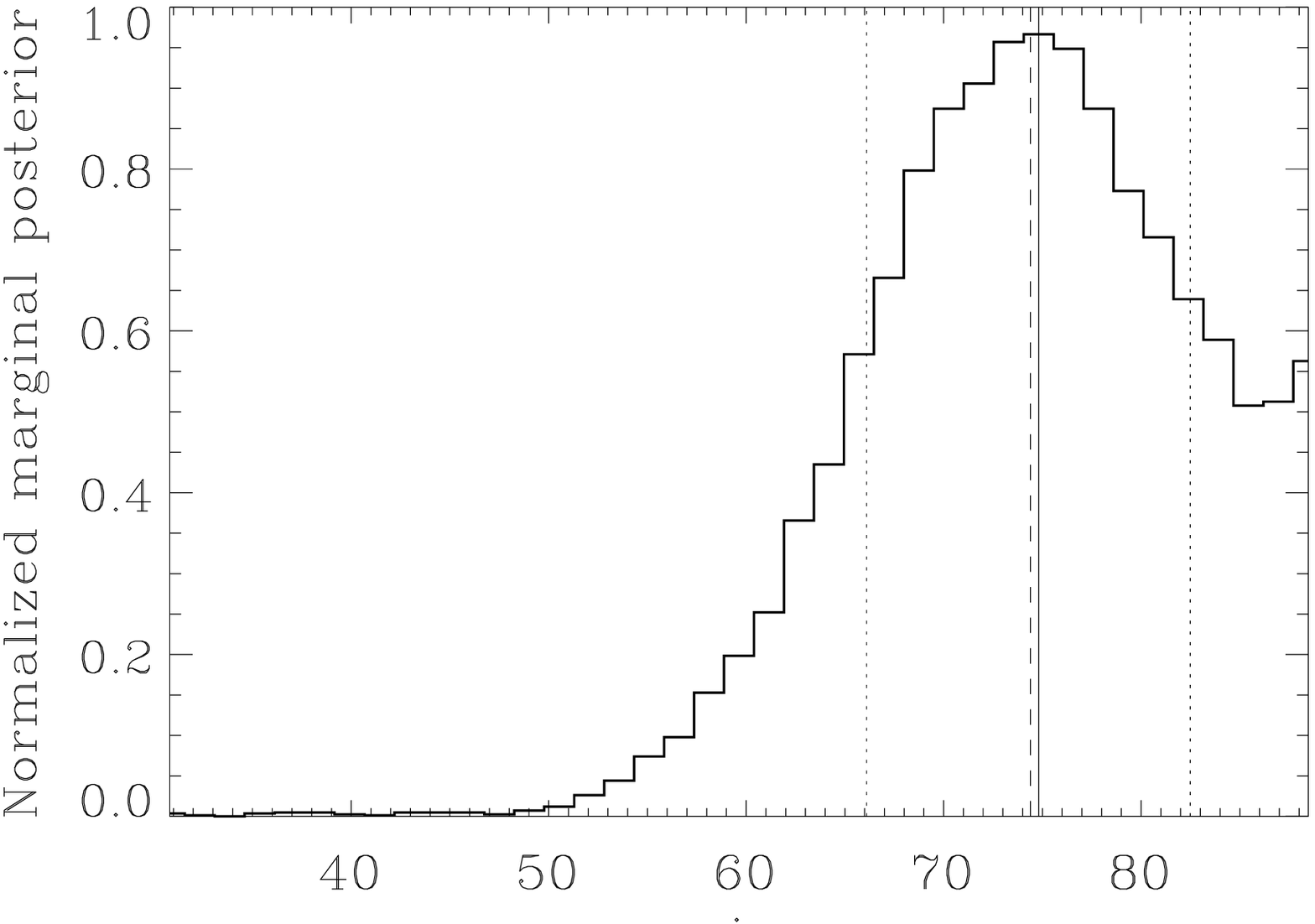}\par} 
\caption{Probability distributions of the free parameters that describe the clumpy models resulting from the fit 
of PKS 0043-42. The vertical shift has been marginalized.
Solid and dashed vertical lines represent the MAP and the median values of each distribution respectively, and the dotted vertical lines indicate 
the 68\% confidence level for each parameter around the median.}
\label{pks0043_distrib}
\end{figure*}

\begin{figure}
\centering
\includegraphics[width=9cm]{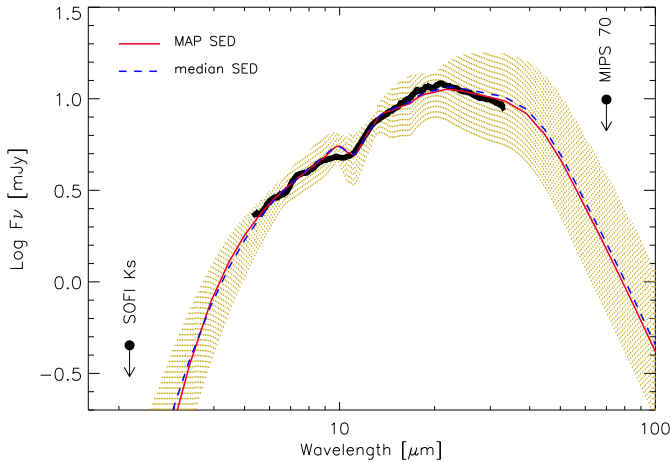}
\caption{Fit of the observed IRS spectrum (thick solid line spanning 5-30 \micron) and the SOFI Ks and MIPS 70 \micron~data.
Dashed and solid lines correspond to the median and MAP models respectively. 
Shaded region indicates the range of models compatible with all the possible combinations of the parameters
within their 68\% confidence intervals around the medians.}
\label{pks0043_fit}
\end{figure}

The result of the SED fitting are the posterior/probability distributions for the six free 
parameters that describe the models (see Table \ref{parametros} for a description of the parameters) 
and the vertical shift (Figure \ref{pks0043_distrib}). 
In addition, we can translate these results into corresponding spectra 
(Figure \ref{pks0043_fit}). Thus, the maximum-a-posteriori (MAP) spectral energy distribution (SED) 
represent the ``best fit'' to the data (solid line), i.e. the combination of parameters 
that maximizes the overall probability/posterior distribution.
The dashed line is the SED obtained using the median
value of the probability distribution of each parameter, and the shaded region in Figure \ref{pks0043_fit}  
indicates the range of models compatible with all the possible combinations of parameters
within their 68\% confidence intervals around the medians.
Figure \ref{pks0043_fit} shows how well the clumpy models fit the observational data. Both the
20 \micron~turnover and the silicate feature in weak absorption are reproduced by the MAP and the median models.

\begin{table*}
\centering
\begin{tabular}{lclll}
\hline
\hline
Parameter & Abbreviation & Interval & \multicolumn{2}{c}{Fitting results} \\
 & & & Median & Mode \\
\hline
Width of the angular distribution of clouds            & $\sigma$        & [15\degr, 75\degr] &  60\degr$\pm^{6\degr}_{13\degr}$ &  68\degr \\
Radial extent of the torus ($R_{o}/R_{d}$)             & $Y$             & [5, 30]            &  12$\pm$1                        &  12      \\
Number of clouds along the radial equatorial direction & $N_0$           & [1, 15]            &   5$\pm$1                        &   5      \\
Power-law index of the radial density profile          & $q$             & [0, 3]             &  0.3$\pm^{0.3}_{0.2}$            &  0.3     \\
Inclination angle of the torus                         & $i$             & [0\degr, 90\degr]  &  74\degr$\pm$8\degr              &  75\degr \\
Optical depth per single cloud                         & $\tau_{V}$      & [10, 200]          &  18$\pm^{5}_{3}$                 &  18      \\ 
\hline
A$_V$ produced by the torus along the LOS              & $A_{V}^{LOS}$   & \dots              &  84$\pm^{11}_{7}$  mag           &  79 mag  \\
\hline      
\end{tabular}						 
\caption{Clumpy model parameters. Columns 1 and 2 give the parameter description and abbreviation used in the text. Column 3 indicates the 
input ranges considered for the fit (i.e., the lower and upper limits of the uniform priors). Finally, columns 4 and 5 list the resulting 
median and mode values of resulting probability distributions resulting from the fit (Figure \ref{pks0043_distrib}).}
\label{parametros}
\end{table*}

\section{Results}
\label{results}

The more information provided by the IR SED, the better the probability distributions are constrained. 
From Figure \ref{pks0043_distrib} it is clear that the IR data of PKS 0043-42 constrain the 
six parameters well (two last columns of Table \ref{parametros}). 
Thus, the IR emission of this radio galaxy can be reproduced by a broad clumpy 
torus (torus width of $\sigma$=60\degr) with an average number of clouds N$_0$=5 along the radial equatorial direction, 
a radial extent $Y$=12, and close to edge-on inclination ($i$=74\degr). The optical extinction produced by the 
torus along the line-of-sight (LOS) is $A_{V}^{LOS} = 1.086~N_0~\tau_{V}~e^{(-(i-90)^{2}/\sigma^{2})}$ = 84 mag.

The clumpy model fits yield the intrinsic bolometric luminosity of 
the AGN by means of the vertical shift applied to match the observational data points: 
L$_{bol}^{MIR}$ = 1.6$\pm^{0.2}_{0.1}\times$10$^{44}~erg~s^{-1}$.
This value can be directly compared with the
unabsorbed 2-10 keV bolometric luminosity measured from Chandra hard X-ray observations
by using a bolometric correction factor of 20 \citep{Elvis94}:
L$_{bol}^X$ = 3.7$\times$10$^{44}~erg~s^{-1}$ (see Section \ref{chandra}). 
Thus, we find L$_{bol}^X$ $\sim$ 2$\times$L$_{bol}^{MIR}$, which is consistent within the uncertainties. 
In fact, these luminosity values are typical of SLRGs \citep{Hardcastle09}, 
and both the MIR (24 and 70 \micron) and [O III]$\lambda$5007 \AA~luminosities reported 
in \citet{Dicken09} for PKS 0043-42 are at the upper end of the WLRG range in the 2Jy sample.

The outer size of the torus ($R_{o} = Y R_{d}$) scales with the AGN bolometric luminosity, 
so assuming a dust sublimation temperature of 1500 K, $R_o= 0.4~Y~(L_{bol}^{MIR}/10^{45})^{0.5}$ pc. 
Considering the median value of $Y$=12 determined from our fit, we
find that the torus in PKS 0043-42 has an outer extent of R$_o\sim$ 2 pc. This value is consistent with 
interferometric observations of nearby AGN, which indicate that the torus emission 
only extends out to 1-2 pc (see e.g. \citealt{Jaffe04,Tristram07}).



Using the A$_V^{LOS}$ value reported in Table \ref{parametros} we can derive the column density 
using the Galactic dust-to-gas ratio (N$_H^{LOS}=1.9\times10^{21}~A_V^{LOS}$; \citealt{Bohlin78}). This gives a value of
N$_H^{LOS}=1.6\pm^{0.2}_{0.1}\times10^{23}~cm^{-2}$, which is similar to that measured from the Chandra X-ray observations
(N$_H^{X-rays}=1.2\pm^{0.5}_{0.4}\times10^{23}~cm^{-2}$; Section \ref{chandra}). 
This indicates that the columns of material implied in the X-ray 
absorption are comparable to those inferred from our IR data. Thus, 
dust-free gas absorption (i.e. gas located inside R$_d$) does not appear to be important here.
This result contrasts with the findings for Seyfert galaxies and QSOs reported by \citet{Maiolino01}, who claimed
that the absorption of the nuclear region in these objects, as derived from optical and IR broad lines, 
is generally much lower than that obtained from the gaseous column density deduced from the X-rays, if a standard
Galactic dust-to-gas ratio is assumed.

\section{Discussion and conclusions}

Based on our analysis, we conclude that the MIR emission of PKS 0043-42 can be accounted for by 
a clumpy dusty torus with the characteristics shown in Table \ref{parametros}. The derived model parameters
are compatible with those obtained for Type-2 AGN of similar AGN bolometric luminosities (see e.g. \citealt{Ramos09}). 
The presence of a torus is also confirmed by hard X-ray Chandra observations, which show clear evidence for a heavily absorbed 
AGN nucleus and a strong iron K-alpha line, which 
are commonly seen in SLRGs but not in WLRGs (Section \ref{chandra}). Indeed, its AGN bolometric luminosity
(L$_{bol}^{X}$ = 3.7$\times$10$^{44}~erg~s^{-1}$) indicates the presence of a moderately luminous hidden AGN.
It has been proposed that the properties of WLRGs, whose spectra lack prominent emission lines, are the result
of the Bondi accretion of hot gas, which prevents the formation of the classical AGN structures such as the torus or the 
BLR. However, the presence of a compact dusty torus and accretion disk signatures in PKS 0043-42 provides evidence that this WLRG is fuelled 
by cold, rather than hot gas accretion.

One question that arises from our result is: if there is a moderately luminous 
AGN heating a nearly edge-on dusty torus, why are we not detecting the strong, high equivalent width emission 
lines characteristic of NLRGs at either optical or MIR wavelengths? It is important to emphasise that, 
although PKS 0043-42 falls close to the correlation between MIR and [O III] luminosity for 2Jy radio galaxies, 
its [O III] luminosity is a factor of ten lower than that of SLRGs at similar redshifts and radio powers \citep{Tadhunter98},
leading to the low equivalent width of the emission lines at optical wavelengths.
On the other hand, 
in the case of the MIR spectrum, the non-linearity of the correlation between emission line and MIR luminosity 
\citep{Dicken09,Dicken10} will naturally lead to the objects at the lower luminosity end of the correlation 
having smaller emission line equivalent widths than their higher luminosity counterparts. In the case of PKS 0043-42, 
this effect is exacerbated by the fact that it lies slightly above ($\sim$0.3 dex, or a factor of two) the 
$L_{24\mu m}$ vs. $L_{[O III]}$ correlation, potentially making the MIR fine structure lines even harder to detect 
against the thermal continuum.

Note that the position of PKS 0043-42 above the $L_{24\mu m}$ vs. $L_{[O III]}$ correlation 
can be explained if the dusty torus structure emitting the MIR continuum has a relatively large covering factor 
compared with the NLR (see discussion in \citealt{Dicken09}). Alternatively, if the illuminating AGN is 
highly variable and has recently ``switched on'', the MIR continuum will be artificially enhanced relative 
to the emission from the NLR, because the compact nuclear dust structure (r$\sim$2 pc according to our modelling) 
will respond more quickly to the change in AGN illumination than the larger-scale NLR (r$\sim$0.1-1 kpc).

It is also interesting to consider the relationship between PKS 0043-42 and the other WLRGs in the 2Jy sample. 
Along with the marked differences in the shape of its MIR continuum spectrum demonstrated in Figure 
\ref{pks0043_spitzer} and discussed in Section \ref{observations}, PKS 0043-42 is 2 -- 10 times more 
luminous at 24 \micron~than the other WLRGs at $z < 0.2$ \citep{Dicken09}. Moreover, for the majority of 
these low redshift WLRGs, the overall MIR continuum shape and extrapolation of the radio core contiuum 
to MIR wavelengths suggest that a substantial fraction of their MIR continuum represents either synchrotron 
\citep{Dicken08} or thermal dust emission associated with starburst components \citep{Leipski09}. 
Therefore, the position of most of the WLRGs close to the $L_{24\mu m}$ vs. $L_{[O III]}$ correlation derived for the 2Jy sample 
as a whole (see Figure 6 in \citealt{Dicken09}) may be misleading, since they would lie below the correlation
if the synchrotron and the starburst contributions were subtracted.
Indeed, there is no evidence for warm dust emission from a torus in the spectra of 
{\it any} of the $z < 0.2$ WLRGs in the 2Jy sample apart from PKS 0043-42. If present, the torus component 
is likely to be sustantially weaker in most WLRGs than it is in PKS 0043-42.

Overall, while PKS 0043-42 is likely to represent a low luminosity counterpart of the SLRGs in terms of its 
nuclear structures and fuelling via cold accretion, such continuity in physical processes is unlikely to 
hold for many low-z WLRGs which show no evidence for warm dust emission associated with the torus.
Therefore our results provide strong evidence that not all WLRGs are fuelled via the same mechanism.

\section*{Acknowledgments}

CRA ackowledges financial support from STFC (ST/G001758/1). 
C.R.A. acknowledges the Spanish Ministry of Science and Innovation (MICINN) through project
Consolider-Ingenio 2010 Program grant CSD2006-00070: First Science with the GTC 
(http://www.iac.es/consolider-ingenio-gtc/).
AAR acknowledges the Spanish Ministry of Science and Innovation through projects AYA2010-18029 
(Solar Magnetism and Astrophysical Spectropolarimetry).
KJI is supported through the Emmy Noether
programme of the German Science Foundation (DFG).
MJH thanks the Royal Society for a fellowship. BM thanks the University of Hertfordshire for a
research studentship.
This work is based on observations made with the Spitzer Space Telescope, 
which is operated by the Jet Propulsion Laboratory, California Institute of 
Technology under a contract with NASA. 
This research has made use of the NASA/IPAC Extragalactic Database (NED) which is 
operated by the Jet Propulsion Laboratory, California Institute of Technology, under 
contract with the National Aeronautics and Space Administration.
CRA acknowledges Carlos Gonz\'{a}lez Fern\' andez and Raffaella Morganti for their valuable help.


\label{lastpage}

\end{document}